\documentclass[a4paper,11pt]{article}
\pdfoutput=1 

\usepackage{jheppub} 
\usepackage{subfigure}
\usepackage{booktabs}
\usepackage{lineno}

\usepackage[T1]{fontenc} 

\def\TeV{\ifmmode {\mathrm{\ Me\kern -0.1em V}}\else \textrm{Te\kern -0.1em V}\fi}%
\def\GeV{\ifmmode {\mathrm{\ Ge\kern -0.1em V}}\else \textrm{Ge\kern -0.1em V}\fi}%
\def\MeV{\ifmmode {\mathrm{\ Me\kern -0.1em V}}\else \textrm{Me\kern -0.1em V}\fi}%

\title{\boldmath Determination of the muonic branching ratio of the $W$ boson and its total width via cross-section measurements at the Tevatron and LHC}

\author[a]{Stefano Camarda}
\author[b]{Jakub Cuth}
\author[b]{Matthias Schott}
\affiliation[a]{CERN, Geneva, Switzerland}
\affiliation[b]{Johannes Gutenberg-University, Mainz, Germany}

\emailAdd{matthias.schott@cern.ch}
\emailAdd{stefano.camarda@cern.ch}

\arxivnumber{1607.05084}

\abstract{
The total $W$-boson decay width $\Gamma_W$ is an important observable
which allows testing of the standard model. The current world average
value is based on direct measurements of final state kinematic
properties of $W$-boson decays, and has a relative uncertainty of 2\%. The
indirect determination of $\Gamma_W$ via the cross-section
measurements of vector-boson production can lead to a similar
accuracy. The same methodology leads also to a determination of the
leptonic branching ratio. This approach has been successfully pursued
by the CDF and D0 experiments at the Tevatron collider, as well as by
the CMS collaboration at the LHC. In this paper we present for the
first time a combination of the available measurements at hadron
colliders, accounting for the correlations of the associated
systematic uncertainties. Our combination leads to values of
$\textrm{BR}(W\rightarrow\mu\nu)=(10.72 \pm 0.16)\%$ and $\Gamma_W = 2113 \pm
31$~\MeV, respectively, both compatible with the current world
averages.
}

\begin{document} 
\maketitle
\flushbottom

\section{\label{sec:Intro}Introduction}	

Precise measurements of the $W$-boson properties, such as its mass $m_W$
and its decay width $\Gamma_W$, allows testing of the standard model
of particle physics. As a matter of fact, the relation between the
$W$-boson mass, $m_W$, the top-quark mass, $m_t$, and the Higgs-boson
mass, $m_H$, via loop corrections, allowed a prediction of the mass of
the Higgs boson with an uncertainty smaller than $25$~\GeV. Models
beyond the standard model could alter the relation between $m_W$ and
$m_H$, since new particles can appear in virtual loops. Similarly, the
total decay width of the $W$ boson can be altered by new
particles.

Within the standard model, the total decay width of the $W$
boson is predicted to be equal to the sum of the partial widths over
three generations of lepton doublets and two generations of quark
doublets. The partial widths are expressed as

\begin{equation}
\label{eqnwidth}
\Gamma_{W\rightarrow f\bar f'} = \frac{|M_{f\bar f'}|^2 \cdot N_C \cdot G_F \cdot m_W^3}{6\pi\sqrt{2}} (1+\delta_f^{\textrm{rad}}(m_t, m_H, ...)),
\end{equation} 
\noindent where $M_{f\bar f'}=1$, $N_C=1$ for leptonic decays, $M_{f\bar f'}$
corresponds to the CKM matrix elements, and $N_C = 3 \cdot (1+\alpha_s(m_W)/\pi + ... )$
is the colour factor for the the quark-decay
modes~\cite{Rosner:1993rj}. Radiative corrections are represented
by $\delta_\ell^{\textrm{rad}}\approx 0.34\%$ for leptons and
$\delta_q^{\textrm{rad}}\approx 0.40\%$ for
quarks~\cite{Renton:2008ub}, which are small in the Standard Model
(SM) since a large part of the corrections is absorbed in the measured
values of $G_F=1.1663787(8)\cdot 10^{-5}$~\GeV$^{-2}$ and
$m_W=80.385\pm0.015$~\GeV. New particle candidates that couple to the
$W$ boson and are lighter than $m_W$, would therefore open a new decay
channel and alter $\Gamma_W$. One very prominent example is
supersymmetric models in which the $W$ boson can decay to the lightest
super-partner of the charged gauge bosons and the lightest
super-partner of the neutral gauge bosons. Hence a precise measurement
of $\Gamma_W$ might reveal physics beyond the standard model.
In addition, assuming SM relations, the dependence of the partial and
total widths of the $W$ boson on the strong-coupling constant allows
to determine the value of $\alpha_s$ from the hadronic and leptonic
branching ratios of the $W$ boson~\cite{d'Enterria:2016ujp}.

The total width of the $W$ boson can be measured directly by kinematic
fits to the measured decay lepton spectra, such as the transverse
momentum of the charged lepton decay $p_T$ or the high-mass tail of
the transverse mass $m_T$ as was performed at CDF and
D0~\cite{Aaltonen:2007ai,Abazov:2009vs,TEW:2010aj}, or via fits to the
invariant mass distributions in the $qqqq$ and $qql\nu$ final states
as was done at the LEP experiments~\cite{Schael:2006mz}. A combination
of these direct results, based on kinematic measurements, leads to
$\Gamma_W = 2085 \pm 42$~\MeV, which is currently used as world
average value~\cite{Agashe:2014kda}.

An independent determination of the $W$-boson width is based on the
measurement of the ratio of cross sections of $W$- and $Z$-boson
production in hadron collisions, defined as

\begin{equation}
\nonumber R = \frac{\sigma(pp' \rightarrow W + X) \cdot \textrm{BR}(W\rightarrow \ell\nu)}{ \sigma(pp' \rightarrow Z + X) \cdot \textrm{BR}(Z\rightarrow \ell\ell) },
\end{equation}
\noindent where $\textrm{BR}(V\rightarrow \ell\ell') = \Gamma_{V\rightarrow
  \ell\ell'} / \Gamma_V$ denotes the leptonic branching ratio of the
vector-boson ($V=W,Z$) decays. The ratio $R$ can be written as

\begin{equation}
\label{eqn:r}
\nonumber R = \frac{\sigma_W}{\sigma_Z} \cdot \frac{\Gamma_{W\rightarrow \ell\nu}}{\Gamma_{W}} \cdot \frac{\Gamma_Z}{\Gamma_{Z\rightarrow \ell\ell}},
\end{equation}
\noindent where the total cross-section ratio $\sigma_W/\sigma_Z$ is
known theoretically to high accuracy~\cite{Catani:2009sm}. The ratio
$\sigma_{Z\rightarrow \ell\ell} / \sigma_{Z}$ was precisely measured
by the LEP experiments and therefore the leptonic branching ratio of
the $W$ boson, $\textrm{BR}(W \rightarrow \ell \nu) =
\Gamma_{W\rightarrow \ell\nu}/\Gamma_{W}$, can be inferred
from the measurement of $R$. The advantage of extracting $\textrm{BR}(W
\rightarrow \ell \nu)$ from the cross-section ratio $R$ lies in
the fact that many experimental systematic uncertainties of each
vector-boson cross-section measurement, such as the uncertainty on the
integrated luminosity, are highly correlated and cancel in the
ratio. The leptonic width of the $W$ boson in the SM can be predicted
by eq.~\ref{eqnwidth} and is $\Gamma(W\rightarrow \ell\nu) =
226.5\pm 0.1$~\MeV.\footnote{Taken from ref.~\cite{Renton:2008ub}, with
  updated values of $m_W$ and $G_F$, and $\alpha_s(m_W)$.} The
dominant uncertainty is due to the accuracy of $m_W$. Using this
value, the total width of the $W$ boson can be extracted by a
measurement of the leptonic branching ratio. This approach for the
determination of the $W$-boson width was already pursued by several
experiments, in particular CDF~\cite{CDFWZ}, D0~\cite{D0WZ}, and
CMS~\cite{CMSWZ1,CMSWZ2}, leading to measurements of $\Gamma_W$
which have an accuracy comparable to the current world average.

In this paper we present a procedure for a first combination of the
individual measurements of the muonic branching ratio of the $W$ boson
and of $\Gamma_W$, accounting for the correlations of the individual
systematic uncertainties. We have chosen to focus on the muon decay
channel, as it has smaller experimental uncertainties.

The paper is structured as follows: we introduce the basic methodology
in section~\ref{sec:Method} and discuss the selected measurements for
the combination in section~\ref{sec:Measurements}, where we also
derive the corresponding fiducial cross-section ratios. The
theoretical predictions of the cross-section ratios are discussed in
section~\ref{sec:theo} and the final extraction and combination of
$\Gamma_W$ for the different experiments is presented in
section~\ref{sec:GammaW}. The paper concludes with a brief summary and
a discussion of the consistency of the results with the direct
measurements and with the global electroweak fit in
section~\ref{sec:sum}.

\section{\label{sec:Method}Methodology}

The production cross section of $W$ and $Z$ bosons in hadron
collisions is described by the Drell-Yan process~\cite{Drell:1970wh}
and can be experimentally defined as

\begin{equation}
\nonumber \sigma_{\textrm{incl}}(pp \rightarrow V + X \rightarrow \ell\ell') = \frac{N_{\textrm{Cand}} - N_{\textrm{Bkg}}}{\epsilon \cdot \int L dt} = \frac{N_{\textrm{Cand}} - N_{\textrm{Bkg}}}{C\cdot A \cdot \int L dt},
\end{equation}
\noindent where $N_{\textrm{Cand}}$ and $N_{\textrm{Bkg}}$ are the
number of vector-boson candidates and the expected background events,
respectively, and $\int L dt$ is the integrated luminosity of the
corresponding data sample. The factor $\epsilon$ is the efficiency of
the signal events passing the signal selection criteria, which is
typically estimated with simulated samples of the signal process, and
corrected for differences in the detector response between data and MC
simulation. The efficiency correction $\epsilon$ can be decomposed as
the product of a fiducial acceptance, $A$, and a detector-induced
correction factor, $C$, i.e. $\epsilon = A \cdot C$. The fiducial
acceptance is the ratio of the number of events that pass the
geometrical and kinematic requirements in the analysis at generator level over
the total number of generated events in a simulated sample of signal
process. The advantage of this decomposition is the separation to a
large extent of detector and analysis related uncertainties, which
enter the factor $C$, while all model and theoretical uncertainties,
such as QCD scales and parton density function (PDF) uncertainties,
enter $A$. The fiducial production cross section
$\sigma_{\textrm{fid}}$ within the detector acceptance volume defined
by A, is barely affected by model uncertainties, and is related to the
fully inclusive cross section by $\sigma_{\textrm{fid}} = \sigma_{\textrm{incl}}\cdot A$.

The strategy for the combination of several indirect
$\textrm{BR}(W\rightarrow \mu \nu)$ and $\Gamma_W$ measurements from
various experiments is therefore based on the measured fiducial
cross-section ratio

\begin{equation}
\label{eqn:rfid}
\nonumber R_{\textrm{fid}} =  \frac{\sigma_{\textrm{fid}}(pp' \rightarrow W + X) \cdot \textrm{BR}(W\rightarrow \ell\nu)}{ \sigma_{\textrm{fid}}(pp' \rightarrow Z + X) \cdot \textrm{BR}(Z\rightarrow \ell\ell) },
\end{equation}
\noindent which has only negligible model uncertainties and
uncorrelated experimental uncertainties between the different
experiments. The fiducial ratio $R_{\textrm{fid}}$ can be related to
the inclusive ratio $R$, by

\begin{equation}
\label{eqn:fid}
\nonumber R = \Big(\frac{A_W}{A_Z}\Big)^{-1} \cdot R_{\textrm{fid}},
\end{equation}
\noindent where the $A_W$ and $A_Z$ are the acceptance correction
factors for the $Z$- and $W$-boson analyses, respectively. Some
published results only present a value for the inclusive cross-section
ratio $R$, but do not publish a value for $R_{\textrm{fid}}$. In these
cases, we have used the fiducial volume definition, the PDF set, and
the MC generator of the corresponding analysis that were used to
extract the acceptance ratio $A_W/A_Z$, in order to reconstruct the
value of $R_{\textrm{fid}}$. The uncertainty on the extrapolated
values of $R_{\textrm{fid}}$ is estimated by subtracting the published
model and PDF uncertainties from the total uncertainty on $R$.

Once the fiducial ratios are determined for each measurement, we can
coherently predict the acceptance correction ratios $A_W/A_Z$ and the
inclusive fiducial cross-section ratios $\sigma_W / \sigma_Z$, and
extract the corresponding branching ratio $\textrm{BR}(W\rightarrow \mu
\nu)$ and decay width $\Gamma_W$ from the measurements of
$R_{\textrm{fid}}$. Each model variation, e.g. one particular
eigenvector variation of a given PDF set, leads to new predictions of
$A_W/A_Z$ and $\sigma_W / \sigma_Z$, thus also to new determined
values of $\textrm{BR}(W\rightarrow \mu \nu)$ and $\Gamma_W$ for each
experiment. The measurements are combined treating the experimental
uncertainties as uncorrelated, and the PDF and model uncertainties
with a correlation model based on a common baseline for the
theoretical predictions.

\section{\label{sec:Measurements}Measurements used for the combination}

One of the first precise measurements of the $\sigma_W / \sigma_Z$
cross-section ratio was published by the D0 collaboration in proton
anti-proton collisions at a centre-of-mass energy of
$\sqrt{s}=1.8$~\TeV~\cite{D0WZ}. However, this measurement was
performed only in the electron decay channel and hence is not used for
this combination. The most precise measurement at the Tevatron
collider was performed by the CDF collaboration at
$\sqrt{s}=1.96$~\TeV~\cite{CDFWZ}, using the electron and muon decay
channels. Only the inclusive cross-section ratio was published
(table~\ref{tab:ProdFigDef}), but the clear definition of the fiducial
volume, as reported in the paper, allows the extrapolation of the
value of $R_{\textrm{fid}}$. The extrapolation factor $A_W/A_Z$ is
estimated using the \textsc{Pythia} 6.2~\cite{Sjostrand:2001yu} generator with
the CTEQ5L PDF set~\cite{Lai:1999wy}.

Several measurements of $R$ were performed at the LHC by the CMS and
ATLAS collaborations in proton-proton collisions at $\sqrt{s}=7$~\TeV{}
and $\sqrt{s}=8$~\TeV~\cite{ATLASWZ,CMSWZ1,CMSWZ2}, which are all used
for the combination. In contrast to the Tevatron experiments, fiducial
ratios together with a fiducial volume definition have also been
published by the ATLAS and CMS experiments, as summarised in
table~\ref{tab:ProdFigDef}. Hence no additional extrapolation to
$R_{\textrm{fid}}$ is performed for these measurements.

\begin{table}[t]
\centering
\begin{small}
\begin{tabular}{c|c|c|c|c}
\toprule
Experiment	&	Collider			& Fiducial							& $R$		& $R_{\textrm{fid}}$		\\
			&					& volume	definition					& (published)		& 				\\		
\midrule
CDF			&	$p\bar p$, 		& $p_T^\mu>20\,\GeV$ for $|\eta|<1.0$	& 10.93 			& 7.46			\\
\cite{CDFWZ}	&	$\sqrt{s}$= 1.96~\TeV	& $Z$ : $66<m_{ee}<116\,\GeV$		& $\pm$0.27 (stat) 	& $\pm$0.18 (stat)	\\
			&				 	& W : $p_T^\nu>20\,\GeV$			& $\pm$0.18 (sys) 	& $\pm$0.12 (sys)	\\
			&					& 								& 				& (extrapolated)	\\

\midrule
ATLAS		&	$pp$, 			& $p_T^\ell>20\,\GeV$ 					& 10.91 			& 10.85			\\
\cite{ATLASWZ}&	$\sqrt{s}$= 7~\TeV	& $|\eta_\ell|<2.5$					& $\pm$0.11 (stat) 	& $\pm$0.11 (stat)	\\
			&				 	& $Z$ : $66<m_{ll}<116\,\GeV$			& $\pm$0.17 (sys) 	& $\pm$0.17 (sys)	\\
			&					& W : $p_T^\nu>25\,\GeV, m_T>40\,\GeV$	& 				& (published)		\\
\midrule
CMS			&	$pp$, 			& $Z$: $p_T^\mu>20\,\GeV$ for $|\eta|<2.1$	& 10.52			& 11.95			\\
\cite{CMSWZ1}	&	$\sqrt{s}$= 7~\TeV	& $Z$: $60<m_{ee}<120\,\GeV$			& $\pm$0.09 (stat) 	& $\pm$0.10 (stat)	\\
			&				 	& W : $p_T^\mu>25\,\GeV$ for $|\eta|<2.1$ & $\pm$0.10 (sys) 	& $\pm$0.20 (sys)	\\
			&					& 								& 				& (published)		\\
\midrule
CMS			&	$pp$, 			& $Z$: $p_T^\mu>25\,\GeV$ for $|\eta|<2.1$	& 10.44 			& 13.28			\\
\cite{CMSWZ2}	&	$\sqrt{s}$= 8~\TeV	& $Z$ : $60<m_{ee}<120\,\GeV$			& $\pm$0.14 (stat) 	& $\pm$0.18 (stat)	\\
			&				 	& W : $p_T^\mu>25\,\GeV$ for $|\eta|<2.1$	& $\pm$0.30 (sys) 	& $\pm$0.23 (sys)	\\
			&					& 								& 				& (published)		\\
\bottomrule
\end{tabular}
\end{small}
\caption{The collider beams, the corresponding centre-of-mass energy,
  the fiducial volume definitions, the published inclusive
  cross-section ratio $R$, as well as the fiducial cross-section ratio
  $R_{\textrm{fid}}$ are given for each analysis used for the
  combination. The fiducial ratio was not published for the
  measurements of the CDF experiment and the extrapolated
  value is shown.}
\label{tab:ProdFigDef}
\end{table}

\section{\label{sec:theo}Theoretical predictions and systematic uncertainties}

The total $W$- and $Z$-boson production cross sections and their
ratio, corresponding to the the fiducial volume definitions of
table~\ref{tab:ProdFigDef}, are calculated at next-to-next-to-leading
order in the perturbative expansion of the strong-coupling constant
with \textsc{FEWZ}~\cite{Gavin:2010az} using the MMHT2014 PDF
set~\cite{Harland-Lang:2014zoa}. The calculations are based on the
$G_\mu$ electroweak parameter scheme and the strong-coupling constant at the
$Z$-boson mass is set to $\alpha_s(m_Z) = 0.118$, as used in the
MMHT2014 PDF determination.\footnote{In this study, we used
  $G_F = 1.1663787\cdot 10^{-5}$~\GeV$^{-2}$, $m_W=80385$~\MeV, $m_Z=91187.6$~\MeV,
  and $\Gamma_Z=2495$~\MeV.} The uncertainties of the PDF set are
estimated by a reevaluation of the predicted cross-section ratio for
each error eigenvector within the MMHT2014 PDF set, as well as the
comparison to the central prediction using a second PDF set, which is
chosen to be the CT10~\cite{Gao:2013xoa} in this study. The
uncertainties, at 68\% CL, include contributions from the
strong-coupling constant $\alpha_s$ as well as variations of the
renormalisation scale, $\mu_R$, and factorisation scale, $\mu_F$.

The correct description of the vector-boson transverse momentum,
$p_T(V=W,Z)$, is essential for the estimation of $A_W$ and
$A_Z$. Since fixed order perturbative QCD predictions do not provide a
sufficiently good description of the low $p_T(V=W,Z)$ spectrum, we use
the \textsc{Powheg} MC generator interfaced to \textsc{Pythia8},
henceforth referred to as \textsc{Powheg+Pythia8}, to estimate the central values
for $A_W$ and $A_Z$, using the MMHT2014 PDF set.

The uncertainties due to missing higher order QCD corrections are
estimated by varying the renormalisation and factorisation scales,
$\mu_R$ and $\mu_F$, by a factor of two up and down, as well as by
reevaluating the acceptance factors with $hdamp$ set to $m_V (V=W,Z)$,
instead of the default value $hdamp = \infty$~\cite{Alioli:2008tz}, in the \textsc{Powheg}
generator. The correlation of the $\mu_R$ and $\mu_F$ variations on
the $W$- and $Z$-boson cross sections and acceptances can be treated
according to various prescriptions. In the most conservative approach
they are considered as fully uncorrelated, leading to an uncertainty
of 0.5\% on the predicted inclusive cross-section ratio. The
uncertainty reduces by more than a factor of two when assuming a fully
correlated behaviour. In the following we adopt an intermediate
approach, and assume a correlation of 50\%. In addition to these
uncertainties, the acceptance factor ratio $A_W/A_Z$ is also affected
by other effects, which change the kinematic distribution of the final
states, but has little effect on the inclusive cross sections. In
particular, the uncertainties due to soft non-perturbative effects and
initial-state radiation (ISR), which vary the transverse momentum
spectrum of the vector boson, $p_T(V)$, have to be estimated. To
perform a conservative estimation, we reweight the predicted $p_T(V)$
from \textsc{Powheg+Pythia8} to corresponding predictions of the
\textsc{Resbos} generator~\cite{Balazs:1997xd, Ladinsky:1993zn,
  Guzzi:2013aja}. \textsc{Resbos} is based on a resummed calculation,
which is performed at next-to-next-to-leading logarithmic order and
matched to approximate NNLO perturbative QCD calculations at large
boson momenta. The difference between the nominal
\textsc{Powheg+Pythia8} predictions of $A_W$ and $A_Z$ and the
\textsc{Resbos} reweighted samples, is considered as an ISR and
resummation uncertainty. The corresponding uncertainties on $A_W/A_Z$
vary between $0.1\%$ (ATLAS) and $0.4\%$ (CMS). This difference can be
explained by the larger effect on the $W$-boson selection for CMS, as it
requires only a minimum threshold on the $p_T$ of the decay muons.

Furthermore, NLO electroweak corrections can be comparable in size to
NNLO QCD corrections. We distinguish between the corrections due to
QED final-state radiation (FSR) and loop-induced electroweak
corrections (EWK). The QED FSR related uncertainties are estimated by
comparing \textsc{Sherpa}~\cite{Gleisberg:2008ta} and
\textsc{Pythia8}~\cite{Sjostrand:2007gs}, where the acceptances are
derived for both generators using dressed and bare leptons. The
resulting differences in the predicted acceptance ratios are taken as
the QED FSR model uncertainty and amount to 0.1\%. The uncertainties
due to loop-induced electroweak corrections are taken from
literature~\cite{Balossini:2008cs} and are accounted for by $0.1\%$
variations on $A_W/A_Z$.

We obtain 57 predictions for the cross-section ratios
$\sigma_W/\sigma_Z$ and the acceptance ratios $A_W/A_Z$, accounting
for 50 MMHT PDF eigenvector variations, the central prediction of the
CT10 PDF set, $\mu_R$ and $\mu_F$ scale variations, and variations of
$\alpha_s$. In addition, we have further uncertainties on $A_W/A_Z$
due to ISR/resummation effects, QED FSR and NLO EWK effects. A summary
of the cross-section ratios $\sigma_W/\sigma_Z$ and acceptance ratios
$A_W/A_Z$ for each measurement, including the relevant model
uncertainties, is given in table~\ref{tab:PredAZWZSigmaWZ}. The PDF
uncertainties are evaluated with the Hessian
method~\cite{Pumplin:2001ct}.

The uncertainties due to ISR and resummation, QED FSR and electroweak
corrections, as well as the variations of $\mu_F$ and $\mu_R$, are
symmetrised by taking the average of the positive and negative
variations.

\begin{table}[t]
\centering
\begin{small}
\begin{tabular}{cccccccc}
\toprule
Experiment	& Quantity						& Value	& Scales				& ISR+			& PDF	& QED FSR + 	& Total	\\
			&							&		& ($\mu_R$, $\mu_F$)	& resummation	&		& EWK		&		\\
\midrule
CDF			& $A_W/A_Z$					& 1.884	& 0.007				& 0.003			& 0.006	& 0.004		& 0.011	\\
			& $(\sigma_W/\sigma_Z)_{\textrm{pred}}$	& 3.391	& 0.005				& -				& 0.013	& -			& 0.014	\\
\midrule
ATLAS		& $A_W/A_Z$					& 1.000	& 0.003				& 0.002			& 0.004	& 0.002		& 0.006	\\
			& $(\sigma_W/\sigma_Z)_{\textrm{pred}}$	& 3.395	& 0.012				& -				& 0.020	& -			& 0.023	\\
\midrule
CMS			& $A_W/A_Z$					& 1.135	& 0.003				& 0.006			& 0.005	& 0.002		& 0.008	\\
(7~\TeV)		& $(\sigma_W/\sigma_Z)_{\textrm{pred}}$	& 3.346	& 0.012				& -				& 0.019	& -			& 0.022	\\
\midrule
CMS			& $A_W/A_Z$					& 1.266	& 0.004				& 0.007			& 0.005	& 0.002		& 0.009	\\
(8~\TeV)		& $(\sigma_W/\sigma_Z)_{\textrm{pred}}$	& 3.326	& 0.013				& -				& 0.020	& -			& 0.023	\\
\bottomrule
\end{tabular}
\end{small}
\caption{Predicted acceptance ratios $A_W/A_Z$ for the extrapolation
  from the experimental fiducial region to the full phase space and
  predicted cross-section ratios $\sigma_W/\sigma_Z$ for all
  measurements under consideration. In addition, the uncertainties due
  to initial-state radiation modelling and resummation model (ISR),
  factorisation and renormalisation scales, PDF, QED final state
  radiation uncertainties as well as electroweak corrections are
  given. It should be noted that the invariant mass requirement of the $Z$-boson
  selection is different between the analyses.}
\label{tab:PredAZWZSigmaWZ}
\end{table}

\section{\label{sec:GammaW}Extraction of the $W$-boson width and combination}

The total inclusive cross-section ratio for each experiment is
estimated by combining the central values of $A_W/A_Z$ reported in
table~\ref{tab:PredAZWZSigmaWZ} and the fiducial cross-section ratios
$R_{\textrm{fid}}$ from table~\ref{tab:ProdFigDef}. It should be noted
that these derived values for $R$ will differ from the original
published values, as our baseline prediction for the estimation of
$A_W/A_Z$ differs from the approach followed by each
experiment. Clearly, the advantage of having a common model for the
theoretical predictions lies in the traceability of correlated
systematic uncertainties. The published values of $R$ are compared to
the values obtained using the newly derived acceptance ratios as a first
sanity check of our extrapolation. The derived values agree with the
published values of the experiments within their associated model
uncertainties.

\begin{figure}[t]
\begin{minipage}[hbt]{.49\textwidth}
	\centering
	\includegraphics[width=0.98\textwidth]{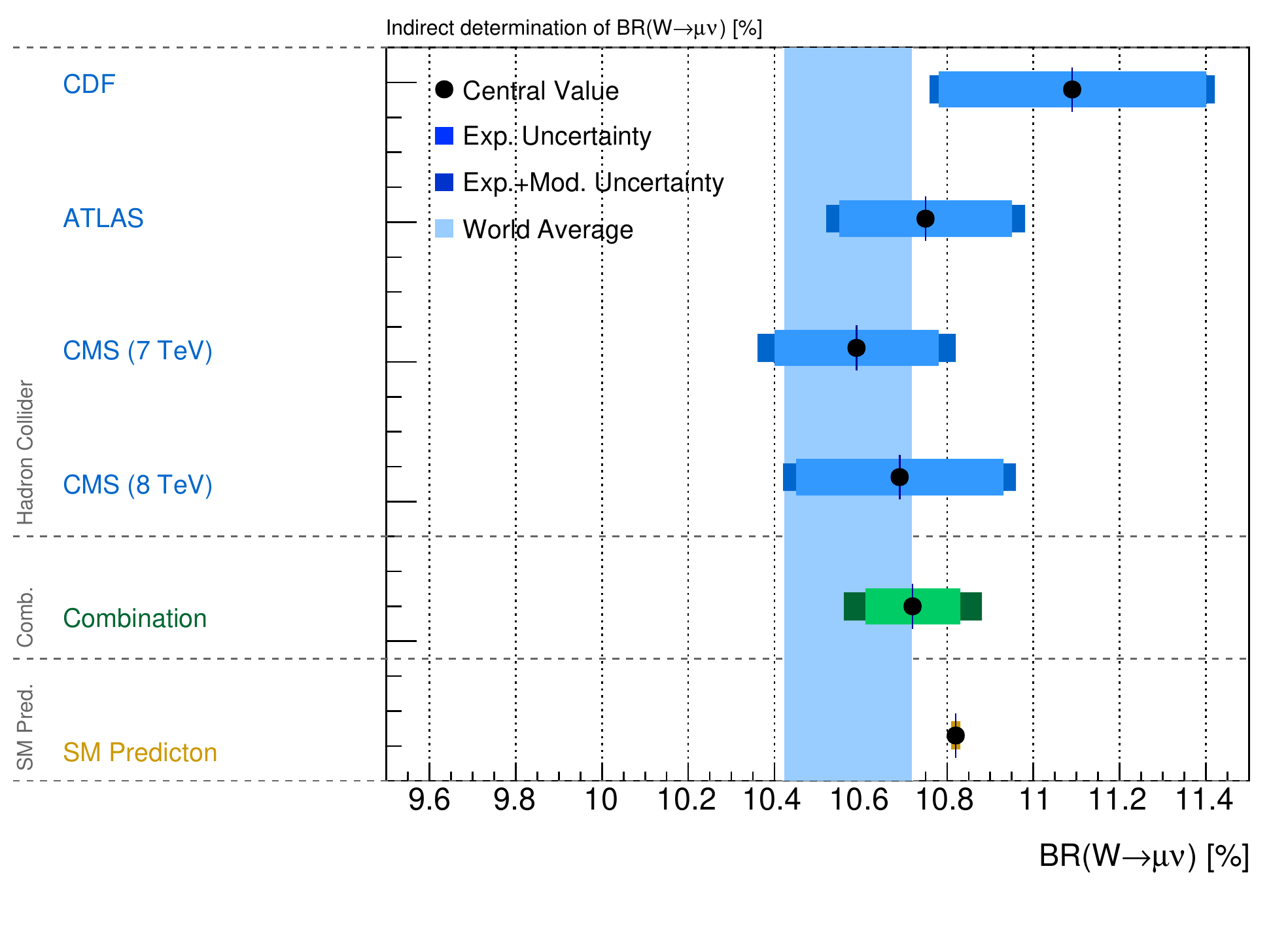}
	\caption{Reestimated $\textrm{BR}(W \rightarrow \mu\nu)$ values
          of the measurements under consideration as well as the
          combined value, the SM prediction, and the current world
          average.\vspace{0.0cm}}
	\label{Fig:CompBR}
\end{minipage}
\hspace{0.2cm}
\begin{minipage}[hbt]{.49\textwidth}
	\centering
	\includegraphics[width=0.98\textwidth]{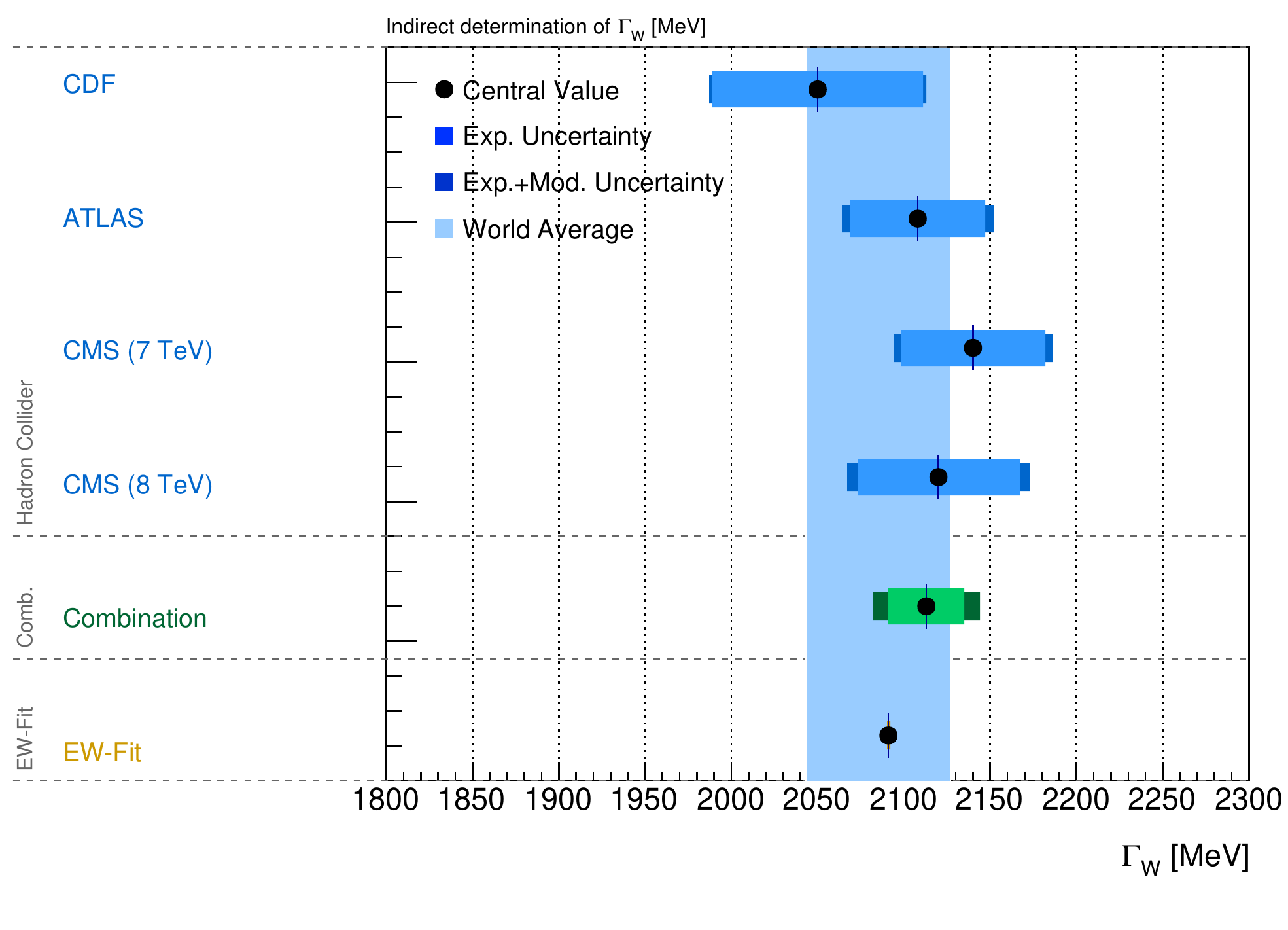}
	\caption{Reestimated $W$-boson width values of the measurements
          under consideration as well as the combined value, the
          result of the global electroweak fit, and the current world
          average.\vspace{0.0cm}}
	\label{Fig:CompGW}
\end{minipage}
\end{figure}

In a second step, the expected leptonic branching ratios can be
rederived for each experiment individually, using the predicted
cross-section ratios, the measured fiducial ratios of the experiments
reported in table~\ref{tab:PredAZWZSigmaWZ}, and the relation

\begin{equation}  
\textrm{BR}(W \rightarrow \mu\nu) = \frac{\Gamma_{W\rightarrow \mu\nu}}{\Gamma_{W}} = R \cdot \frac{\sigma_Z}{\sigma_W} \cdot  \frac{\Gamma_{Z\rightarrow \mu\mu}}{\Gamma_{Z}},
\end{equation}  

\noindent where a leptonic $Z$-boson branching ratio of
$\Gamma_{Z\rightarrow \mu\mu}/ \Gamma_{Z} = 0.033658 \pm
0.000023$~\cite{Agashe:2014kda}
is used. The results are presented in
table~\ref{tab:DerivedValuesBR}. Assuming the validity of the SM, the
partial leptonic $W$-boson width is predicted by
eq.~\ref{eqnwidth}, leading to $\Gamma_{W\rightarrow \mu\nu}=226.5$~\MeV,
where the $(1 + \delta^{\textrm{rad}})$ corrections are taken from
ref.~\cite{Renton:2008ub}. Finally, the total $W$-boson width can be
derived from the measured leptonic branching ratios. The resulting
values for $\Gamma_W = \textrm{BR} \cdot \Gamma_{W \to \mu \nu}$ for each
experimental measurement are illustrated in figure~\ref{Fig:CompGW}
and reported in table~\ref{tab:DerivedValuesGW}, together with the
associated statistical, experimental systematic, and combined model
uncertainties.

\begin{table}[t]
\centering
\begin{small}
\begin{tabular}{ccccccccc}
\toprule
Experiment	& BR	.	& Stat.	& Exp. sys.	& Scales				& ISR+			& PDF	& FSR + 	& Total	\\
			& [\%]	&		&			& ($\mu_R$, $\mu_F$)	& resummation	&		& EWK	&		\\
\midrule
ATLAS		& 10.75	& 0.11	& 0.17		& 0.05				& 0.02			& 0.10	& 0.01	& 0.23	\\
\midrule
CMS (7~\TeV)	& 10.59	& 0.09	& 0.18		& 0.05				& 0.05			& 0.10	& 0.02	& 0.23	\\
\midrule
CMS (8~\TeV)	& 10.69	& 0.14	& 0.19		& 0.06				& 0.05			& 0.11	& 0.02	& 0.27	\\
\midrule
CDF			& 11.06	& 0.27	& 0.18		& 0.04				& 0.02			& 0.07	& 0.02	& 0.33	\\
\midrule
Combined	& 10.72	& 0.07	& 0.09		& 0.04				& 0.02			& 0.10	& 0.02	& 0.16	\\
\bottomrule
\end{tabular}
\end{small}
\caption{Extracted values of $\textrm{BR}(W\rightarrow \mu\nu)$ [\%] for all four measurements and their combination. The associated statistical, experimental, and model uncertainties are also given.}
\label{tab:DerivedValuesBR}
\end{table}

\begin{table}[t]
\centering
\begin{small}
\begin{tabular}{ccccccccc}
\toprule
Experiment	& $\Gamma_W$	& Stat.	& Exp. sys.	& Scales				& ISR+			& PDF	& FSR + 	& Total	\\
			&  [\MeV]			&		&			& ($\mu_R$, $\mu_F$)	& resummation	&		& EWK	&		\\
\midrule
ATLAS		& 2108	& 21		& 33		& 10 			& 3 				& 18 		& 4 		& 44		\\
\midrule
CMS (7~\TeV)	& 2140	& 18		& 36		& 9 			& 11 				& 18 		& 4 		& 46		\\
\midrule
CMS (8~\TeV)	& 2120	& 29		& 37		& 12 			& 11 				& 19 		& 4 		& 53		\\
\midrule
CDF			& 2050	& 51		& 34		& 8 			& 4 				& 13 		& 4 		& 63		\\
\midrule
Combined	& 2113	& 13		& 18		& 8 			& 6 				& 19 		& 4 		& 31		\\
\bottomrule
\end{tabular}
\end{small}
\caption{Extracted values of $\Gamma_W$ [\MeV] for all four
  measurements and their combination. The associated statistical,
  experimental, and model uncertainties are also given.}
\label{tab:DerivedValuesGW}
\end{table}

For the combination, we use the measured inclusive cross-section ratio
$R$ and the corresponding inclusive cross-section prediction for each
model systematic variation, thus leading to new values of $\textrm{BR}(W
\rightarrow \mu\nu)$ and $\Gamma_W$ for each measurement,
respectively. In a second step, we combine the individual measurements
following the BLUE method~\cite{Lyons:1988rp}, again, separately for
all model variations. For the combination of the four experimental
values of $\textrm{BR}(W \rightarrow \mu\nu)$ and $\Gamma_W$, we treat
the statistical and experimental systematic uncertainties as fully
uncorrelated. In a last step, we calculate the difference between the
combined values of $\textrm{BR}(W \rightarrow \ell\nu)$ and $\Gamma_W$
for each model variation and their central combined values, and
evaluate the theoretical and model systematic uncertainties from these
differences, according to standard procedures. The results of the
combination are
\begin{eqnarray}
  \nonumber \textrm{BR}(W \rightarrow \mu\nu) &=& (10.72 \pm 0.07 (\textrm{stat.}) \pm 0.09 (\textrm{exp.syst.})  \pm 0.11 (\textrm{mod.syst.}))\% \\
  \nonumber &=& (10.72 \pm 0.16)\%
\end{eqnarray}
\noindent and 
\begin{eqnarray}
  \nonumber \Gamma_{W} &=& 2113 \pm 13 (\textrm{stat.}) \pm 18 (\textrm{exp.syst.})  \pm 22 (\textrm{mod.syst.}) \,\MeV \\
  \nonumber &=&  2113 \pm 31 \,\MeV.
\end{eqnarray}
The results are shown and compared to the current world
averages and to the SM predictions in figures~\ref{Fig:CompBR}
and~\ref{Fig:CompGW}, respectively. The statistical and systematic
uncertainties of the combined measurement of $\Gamma_W$ are 30\% and
45\% smaller, respectively, compared to the uncertainties of the most
precise single measurement. The model uncertainties on the combined
values does not significantly change and are dominated by PDF
uncertainties.

\section{\label{sec:sum}Summary and interpretation}

In this paper we have presented a combination of measurements of the
muonic branching ratio of the $W$ boson and its total decay width,
extracted from the cross-section ratios of $W$- and $Z$-boson production
from the ATLAS, CMS, and CDF experiments at various centre-of-mass
energies. Special emphasis was drawn to the correct treatment of the
correlations between systematic uncertainties, in particular
uncertainties due to the limited knowledge of the parton distribution
functions and variations of the renormalisation and factorisation
scales. The combination yields $\textrm{BR}(W\rightarrow \mu\nu)^{R} =
(10.72\pm0.16)\%$ and $\Gamma^{R}_W = 2113\pm31$~\MeV. The results
are compatible in value, and similar in accuracy, to the current world
averages $\textrm{BR}(W\rightarrow \mu\nu)^{\textrm{WA}} = (10.57\pm0.15)\%$
and $\Gamma_W^{\textrm{WA}} = 2085\pm42$~\MeV, which are based
solely on direct measurements.

The total width of the $W$ boson is potentially sensitive to new
physics scenarios in the context of the global electroweak
fit~\cite{ALEPH:2010aa}.  The indirect determination via the
electroweak fit yields a value of $\Gamma_W^{EW} =
2091\pm1$~\MeV~\cite{Baak:2014ora}, which is in good agreement with
our combined value. The relation expressed in eq.~\ref{eqnwidth}
shows that $\Gamma_W$ depends, among other SM parameters, on $m_W$,
$\alpha_s$, and $m_H$. However, the small uncertainties on the
determination of $\Gamma_W^{EW}$ indicates that the sensitivity of
$\Gamma_W$ to these parameters of the SM is rather weak.

It should be noted that loop corrections arising from contributions of
new physics to the $W$-boson width would alter the term
$\delta^{\textrm{rad}}$ in eq.~\ref{eqnwidth} independently from
the decay channel. As a consequence, the branching ratio is
insensitive to effects that could appear in the corresponding loop
correction terms. Only new physics effects that directly alter the
leptonic branching ratio can be tested with our combined value
$\textrm{BR}(W\rightarrow \mu\nu)^{R}$. We find a perfect agreement with the SM
prediction $\textrm{BR}(W\rightarrow \mu\nu)^{SM} = (10.83 \pm
0.01)\%$~\cite{Rosner:1993rj, Renton:2008ub}.\footnote{Using
  $\alpha_s(m_W) = 0.1203$, corresponding to the value of $\alpha_s(m_Z)$ used
  for the MMHT2014 PDF set, the SM prediction of $\textrm{BR}(W\rightarrow
  \mu\nu)^{SM}$ changes to $10.82\%$.}

\section*{Acknowledgments}

We would like to thank M. Boonekamp, A. Glazov, K. Moenig, and S. Webb
for the useful comments during the preparation this paper. This work
was partly supported by the Volkswagen Foundation and the German
Research Foundation (DFG).

\bibliography{WBosonWidth}{}

\providecommand{\href}[2]{#2}\begingroup\raggedright\begin{thebibliography}{10}

\bibitem{Rosner:1993rj}
J.~L. Rosner, M.~P. Worah and T.~Takeuchi, \emph{{Oblique corrections to the
  $W$ width}}, \href{http://dx.doi.org/10.1103/PhysRevD.49.1363}{\emph{Phys.
  Rev. D} {\bf 49} (1994) 1363--1369},
  [\href{http://arxiv.org/abs/hep-ph/9309307}{{\tt hep-ph/9309307}}].

\bibitem{Renton:2008ub}
P.~Renton, \emph{{Updated SM calculations of $\sigma_W$ / $\sigma_Z$ at the
  Tevatron and the $W$ boson width}},
  \href{http://arxiv.org/abs/0804.4779}{{\tt 0804.4779}}.

\bibitem{d'Enterria:2016ujp}
D.~d'Enterria and M.~Srebre, \emph{{$\alpha_s$, $\rm V_{cs}$, and CKM unitarity
  test from W decays at NNLO}},  \href{http://arxiv.org/abs/1603.06501}{{\tt
  1603.06501}}.

\bibitem{Aaltonen:2007ai}
{\scshape CDF} collaboration, T.~Aaltonen et~al., \emph{{A Direct measurement
  of the $W$ boson width in $p \bar{p}$ collisions at $\sqrt{s}$ = 1.96-TeV}},
  \href{http://dx.doi.org/10.1103/PhysRevLett.100.071801}{\emph{Phys. Rev.
  Lett.} {\bf 100} (2008) 071801}, [\href{http://arxiv.org/abs/0710.4112}{{\tt
  0710.4112}}].

\bibitem{Abazov:2009vs}
{\scshape D0} collaboration, V.~M. Abazov et~al., \emph{{Direct measurement of
  the W boson width}},
  \href{http://dx.doi.org/10.1103/PhysRevLett.103.231802}{\emph{Phys. Rev.
  Lett.} {\bf 103} (2009) 231802}, [\href{http://arxiv.org/abs/0909.4814}{{\tt
  0909.4814}}].

\bibitem{TEW:2010aj}
T.~E.~W. Group, \emph{{Combination of CDF and D0 Results on the Width of the W
  boson}},  \href{http://arxiv.org/abs/1003.2826}{{\tt 1003.2826}}.

\bibitem{Schael:2006mz}
{\scshape ALEPH} collaboration, S.~Schael et~al., \emph{{Measurement of the $W$
  boson mass and width in $e^{+} e^{-}$ collisions at LEP}},
  \href{http://dx.doi.org/10.1140/epjc/s2006-02576-8}{\emph{Eur. Phys. J.} {\bf
  C47} (2006) 309--335}, [\href{http://arxiv.org/abs/hep-ex/0605011}{{\tt
  hep-ex/0605011}}].

\bibitem{Agashe:2014kda}
{\scshape Particle Data Group} collaboration, K.~Olive et~al., \emph{{Review of
  Particle Physics}},
  \href{http://dx.doi.org/10.1088/1674-1137/38/9/090001}{\emph{Chin. Phys.}
  {\bf C38} (2014) 090001}.

\bibitem{Catani:2009sm}
S.~Catani, L.~Cieri, G.~Ferrera, D.~de~Florian and M.~Grazzini, \emph{{Vector
  boson production at hadron colliders: a fully exclusive QCD calculation at
  NNLO}}, \href{http://dx.doi.org/10.1103/PhysRevLett.103.082001}{\emph{Phys.
  Rev. Lett.} {\bf 103} (2009) 082001},
  [\href{http://arxiv.org/abs/0903.2120}{{\tt 0903.2120}}].

\bibitem{CDFWZ}
{\scshape CDF} collaboration, A.~Abulencia et~al., \emph{{Measurements of
  inclusive W and Z cross sections in p anti-p collisions at $\sqrt{s} =
  1.96$~TeV}}, \href{http://dx.doi.org/10.1088/0954-3899/34/12/001}{\emph{J.
  Phys. G} {\bf 34} (2007) 2457--2544},
  [\href{http://arxiv.org/abs/hep-ex/0508029}{{\tt hep-ex/0508029}}].

\bibitem{D0WZ}
{\scshape D0} collaboration, B.~Abbott et~al., \emph{{Extraction of the width
  of the $W$ boson from measurements of $\sigma(p\bar{p} \to W + X) \times B(W
  \to e \nu)$ and $\sigma(p\bar{p} \to Z + X) \times B(Z \to e e)$ and their
  ratio}}, \href{http://dx.doi.org/10.1103/PhysRevD.61.072001}{\emph{Phys. Rev.
  D} {\bf 61} (2000) 072001}, [\href{http://arxiv.org/abs/hep-ex/9906025}{{\tt
  hep-ex/9906025}}].

\bibitem{CMSWZ1}
{\scshape CMS} collaboration, S.~Chatrchyan et~al., \emph{{Measurement of the
  Inclusive $W$ and $Z$ Production Cross Sections in $pp$ Collisions at
  $\sqrt{s}=7$ TeV}},
  \href{http://dx.doi.org/10.1007/JHEP10(2011)132}{\emph{JHEP} {\bf 10} (2011)
  132}, [\href{http://arxiv.org/abs/1107.4789}{{\tt 1107.4789}}].

\bibitem{CMSWZ2}
{\scshape CMS} collaboration, S.~Chatrchyan et~al., \emph{{Measurement of
  inclusive W and Z boson production cross sections in pp collisions at
  $\sqrt{s}$ = 8 TeV}},
  \href{http://dx.doi.org/10.1103/PhysRevLett.112.191802}{\emph{Phys. Rev.
  Lett.} {\bf 112} (2014) 191802}, [\href{http://arxiv.org/abs/1402.0923}{{\tt
  1402.0923}}].

\bibitem{Drell:1970wh}
S.~D. Drell and T.-M. Yan, \emph{{Massive Lepton Pair Production in
  Hadron-Hadron Collisions at High-Energies}},
  \href{http://dx.doi.org/10.1103/PhysRevLett.25.316}{\emph{Phys. Rev. Lett.}
  {\bf 25} (1970) 316--320}.

\bibitem{Sjostrand:2001yu}
T.~Sjostrand, L.~Lonnblad and S.~Mrenna, \emph{{PYTHIA 6.2: Physics and
  manual}},  \href{http://arxiv.org/abs/hep-ph/0108264}{{\tt hep-ph/0108264}}.

\bibitem{Lai:1999wy}
{\scshape CTEQ} collaboration, H.~Lai et~al., \emph{{Global QCD analysis of
  parton structure of the nucleon: CTEQ5 parton distributions}},
  \href{http://dx.doi.org/10.1007/s100529900196}{\emph{Eur. Phys. J.} {\bf C12}
  (2000) 375--392}, [\href{http://arxiv.org/abs/hep-ph/9903282}{{\tt
  hep-ph/9903282}}].

\bibitem{ATLASWZ}
{\scshape ATLAS} collaboration, G.~Aad et~al., \emph{{Measurement of the
  inclusive $W^\pm$ and Z/gamma cross sections in the electron and muon decay
  channels in $pp$ collisions at $\sqrt{s}=7$ TeV with the ATLAS detector}},
  \href{http://dx.doi.org/10.1103/PhysRevD.85.072004}{\emph{Phys. Rev. D} {\bf
  85} (2012) 072004}, [\href{http://arxiv.org/abs/1109.5141}{{\tt 1109.5141}}].

\bibitem{Gavin:2010az}
R.~Gavin, Y.~Li, F.~Petriello and S.~Quackenbush, \emph{{FEWZ 2.0: A code for
  hadronic Z production at next-to-next-to-leading order}},
  \href{http://dx.doi.org/10.1016/j.cpc.2011.06.008}{\emph{Comput. Phys.
  Commun.} {\bf 182} (2011) 2388--2403},
  [\href{http://arxiv.org/abs/1011.3540}{{\tt 1011.3540}}].

\bibitem{Harland-Lang:2014zoa}
L.~A. Harland-Lang, A.~D. Martin, P.~Motylinski and R.~S. Thorne, \emph{{Parton
  distributions in the LHC era: MMHT 2014 PDFs}},
  \href{http://dx.doi.org/10.1140/epjc/s10052-015-3397-6}{\emph{Eur. Phys. J.}
  {\bf C75} (2015) 204}, [\href{http://arxiv.org/abs/1412.3989}{{\tt
  1412.3989}}].

\bibitem{Gao:2013xoa}
J.~Gao et~al., \emph{{CT10 next-to-next-to-leading order global analysis of
  QCD}}, \href{http://dx.doi.org/10.1103/PhysRevD.89.033009}{\emph{Phys. Rev.
  D} {\bf 89} (2014) 033009}, [\href{http://arxiv.org/abs/1302.6246}{{\tt
  1302.6246}}].

\bibitem{Alioli:2008tz}
S.~Alioli, P.~Nason, C.~Oleari and E.~Re, \emph{{NLO Higgs boson production via
  gluon fusion matched with shower in POWHEG}},
  \href{http://dx.doi.org/10.1088/1126-6708/2009/04/002}{\emph{JHEP} {\bf 04}
  (2009) 002}, [\href{http://arxiv.org/abs/0812.0578}{{\tt 0812.0578}}].

\bibitem{Balazs:1997xd}
C.~Balazs and C.~P. Yuan, \emph{{Soft gluon effects on lepton pairs at hadron
  colliders}}, \href{http://dx.doi.org/10.1103/PhysRevD.56.5558}{\emph{Phys.
  Rev. D} {\bf 56} (1997) 5558--5583},
  [\href{http://arxiv.org/abs/hep-ph/9704258}{{\tt hep-ph/9704258}}].

\bibitem{Ladinsky:1993zn}
G.~A. Ladinsky and C.~P. Yuan, \emph{{The Nonperturbative regime in QCD
  resummation for gauge boson production at hadron colliders}},
  \href{http://dx.doi.org/10.1103/PhysRevD.50.R4239}{\emph{Phys. Rev. D} {\bf
  50} (1994) 4239}, [\href{http://arxiv.org/abs/hep-ph/9311341}{{\tt
  hep-ph/9311341}}].

\bibitem{Guzzi:2013aja}
M.~Guzzi, P.~M. Nadolsky and B.~Wang, \emph{{Nonperturbative contributions to a
  resummed leptonic angular distribution in inclusive neutral vector boson
  production}}, \href{http://dx.doi.org/10.1103/PhysRevD.90.014030}{\emph{Phys.
  Rev. D} {\bf 90} (2014) 014030}, [\href{http://arxiv.org/abs/1309.1393}{{\tt
  1309.1393}}].

\bibitem{Gleisberg:2008ta}
T.~Gleisberg, S.~Hoeche, F.~Krauss, M.~Schonherr, S.~Schumann, F.~Siegert
  et~al., \emph{{Event generation with SHERPA 1.1}},
  \href{http://dx.doi.org/10.1088/1126-6708/2009/02/007}{\emph{JHEP} {\bf 02}
  (2009) 007}, [\href{http://arxiv.org/abs/0811.4622}{{\tt 0811.4622}}].

\bibitem{Sjostrand:2007gs}
T.~Sjostrand, S.~Mrenna and P.~Z. Skands, \emph{{A Brief Introduction to PYTHIA
  8.1}}, \href{http://dx.doi.org/10.1016/j.cpc.2008.01.036}{\emph{Comput. Phys.
  Commun.} {\bf 178} (2008) 852--867},
  [\href{http://arxiv.org/abs/0710.3820}{{\tt 0710.3820}}].

\bibitem{Balossini:2008cs}
G.~Balossini, G.~Montagna, C.~M. Carloni~Calame, M.~Moretti, M.~Treccani,
  O.~Nicrosini et~al., \emph{{Electroweak \& QCD corrections to Drell Yan
  processes}}, {\emph{Acta Phys. Polon.} {\bf B39} (2008) 1675},
  [\href{http://arxiv.org/abs/0805.1129}{{\tt 0805.1129}}].

\bibitem{Pumplin:2001ct}
J.~Pumplin, D.~Stump, R.~Brock, D.~Casey, J.~Huston, J.~Kalk et~al.,
  \emph{{Uncertainties of predictions from parton distribution functions. 2.
  The Hessian method}},
  \href{http://dx.doi.org/10.1103/PhysRevD.65.014013}{\emph{Phys. Rev. D} {\bf
  65} (2001) 014013}, [\href{http://arxiv.org/abs/hep-ph/0101032}{{\tt
  hep-ph/0101032}}].

\bibitem{Lyons:1988rp}
L.~Lyons, D.~Gibaut and P.~Clifford, \emph{{How to Combine Correlated Estimates
  of a Single Physical Quantity}},
  \href{http://dx.doi.org/10.1016/0168-9002(88)90018-6}{\emph{Nucl. Instrum.
  Meth. A} {\bf 270} (1988) 110}.

\bibitem{ALEPH:2010aa}
{\scshape ALEPH, CDF, D0, DELPHI, L3, OPAL, SLD, LEP Electroweak Working Group,
  Tevatron Electroweak Working Group, SLD Electroweak and Heavy Flavour Groups}
  collaboration, L.~E.~W. Group, \emph{{Precision Electroweak Measurements and
  Constraints on the Standard Model}},
  \href{http://arxiv.org/abs/1012.2367}{{\tt 1012.2367}}.

\bibitem{Baak:2014ora}
{\scshape Gfitter Group} collaboration, M.~Baak et~al., \emph{{The global
  electroweak fit at NNLO and prospects for the LHC and ILC}},
  \href{http://dx.doi.org/10.1140/epjc/s10052-014-3046-5}{\emph{Eur. Phys. J.}
  {\bf C74} (2014) 3046}, [\href{http://arxiv.org/abs/1407.3792}{{\tt
  1407.3792}}].

\end{thebibliography}\endgroup
\bibliographystyle{JHEP}

\end{document}